\documentclass[12pt,letterpaper]{article}

\usepackage{graphicx,array}
\usepackage{url}
\usepackage{color}
\usepackage{latexsym}
\usepackage{amsthm}
\usepackage{amsmath}
\usepackage{amssymb}
\usepackage{amsfonts}
\usepackage[numbers,sort&compress]{natbib}
\usepackage{bm}
\usepackage{slashed}
\usepackage{mathrsfs}
\usepackage{enumerate}
\usepackage{tikz}
\usepackage{siunitx}
\usepackage{mdframed}
\usepackage{setspace}  
\usepackage{esvect}

\usepackage[utf8x]{inputenc}%
\usepackage{tcolorbox}%

\usepackage{hyperref} 
\hypersetup{
    colorlinks=true,       
    linkcolor=red,          
    citecolor=blue,        
    filecolor=magenta,      
    urlcolor=blue           
}
\usepackage[all]{hypcap} 

\usepackage{natbib}
\usepackage[normalem]{ulem} 

\newcommand{\nc}{\newcommand}

\nc{\beq}{\begin{equation}}  
\nc{\eeq}{\end{equation}}  
\nc{\beqa}{\begin{eqnarray}}  
\nc{\eeqa}{\end{eqnarray}}  
\nc{\bit}{\begin{itemize}}  
\nc{\eit}{\end{itemize}}  

\def\GeV{\mathrm{GeV}}     

\newcommand{\eg}{{\it e.g.}}
\newcommand{\ie}{{\it i.e.}}

\newcommand{\cO}{\mathcal{O}}

\usepackage{floatrow}
\newfloatcommand{capbtabbox}{table}[][\FBwidth]

\usepackage{blindtext}

\newcommand{\ball}{{\tikz{\shade [ball color=gray] (0,0) circle [radius=2.5pt];}}}

\title{ {\bf Catalyzed Baryogenesis}
\author{\large Yang Bai$^{\,\star}$, Joshua Berger$\,^\diamond$, Mrunal Korwar$^{\,\star}$, and Nicholas Orlofsky$^{\,\dagger}$}
\date{\small \it 
$^\star$Department of Physics, University of Wisconsin-Madison, Madison, WI 53706, USA\\
$^\diamond$Department of Physics, Colorado State University, Fort Collins, Colorado 80523, USA \\
$^\dagger$Department of Physics, Carleton University, Ottawa, ON K1S 5B6, Canada \\
}
}

\begin{document}

\maketitle

\setlength{\parskip}{0.2ex}

\begin{abstract}	
A novel mechanism, ``catalyzed baryogenesis," is proposed to explain the observed baryon asymmetry in our universe. In this mechanism, the motion of a ball-like catalyst provides the necessary out-of-equilibrium condition, its outer wall has CP-violating interactions with the Standard Model particles, and its interior has baryon number violating interactions. We use the electroweak-symmetric ball model as an example of such a catalyst. In this model, electroweak sphalerons inside the ball are active and convert baryons into leptons. The observed baryon number asymmetry can be produced for a light ball mass and a large ball radius. Due to direct detection constraints on relic balls, we consider a scenario in which the balls evaporate, leading to dark radiation at testable levels. 
\end{abstract}

\thispagestyle{empty}  
\newpage  
  
\setcounter{page}{1}  
%
%

\section{Introduction}\label{sec:Introduction}

One of the biggest challenges in particle physics and cosmology is to understand the matter-antimatter asymmetry in our universe. More precisely, baryogenesis models must explain the small (dimensionless) baryon-to-photon ratio: $\eta =(N_{\rm baryon} - N_{\rm antibaryon})/N_\gamma$ in the current universe with $N_{\rm baryon}$, $N_{\rm antibaryon}$, and $N_\gamma$ as the numbers of baryons, anti-baryons, and photons, respectively. The measurement of light element abundances in the intergalactic medium together with the theory of Big Bang nucleosynthesis (BBN) has $\eta = (5.931\pm 0.051) \times 10^{-10}$~\cite{Cooke:2017cwo}, while the spectrum of the temperature fluctuations of 
the cosmic microwave background (CMB) has $\eta = (6.12\pm 0.04) \times 10^{-10}$~\cite{Aghanim:2018eyx,Zyla:2020zbs}. Sakharov proposed three general conditions necessary for any baryogenesis model: i) baryon number violation (BNV), ii) C and CP violation, and iii) departure from thermal equilibrium~\cite{Sakharov:1967dj}. 
There are many mechanisms proposed in the literature, some of which have phenomenologically testable predictions (\eg, electroweak baryogenesis~\cite{Cohen:1993nk}). In this paper, we propose a framework of baryogenesis, ``catalyzed baryogenesis" (CAB), wherein BNV interactions are enhanced in certain regions of space. These regions are similar to ``catalysts'' in chemistry where certain molecules aid the interactions of other molecules. If the catalysts are sufficiently long-lived, they could even be detected as relics making up dark matter today.

\begin{figure}[t]
	\label{fig:catalyze-schematic}
	\begin{center}
		\includegraphics[width=0.7\textwidth]{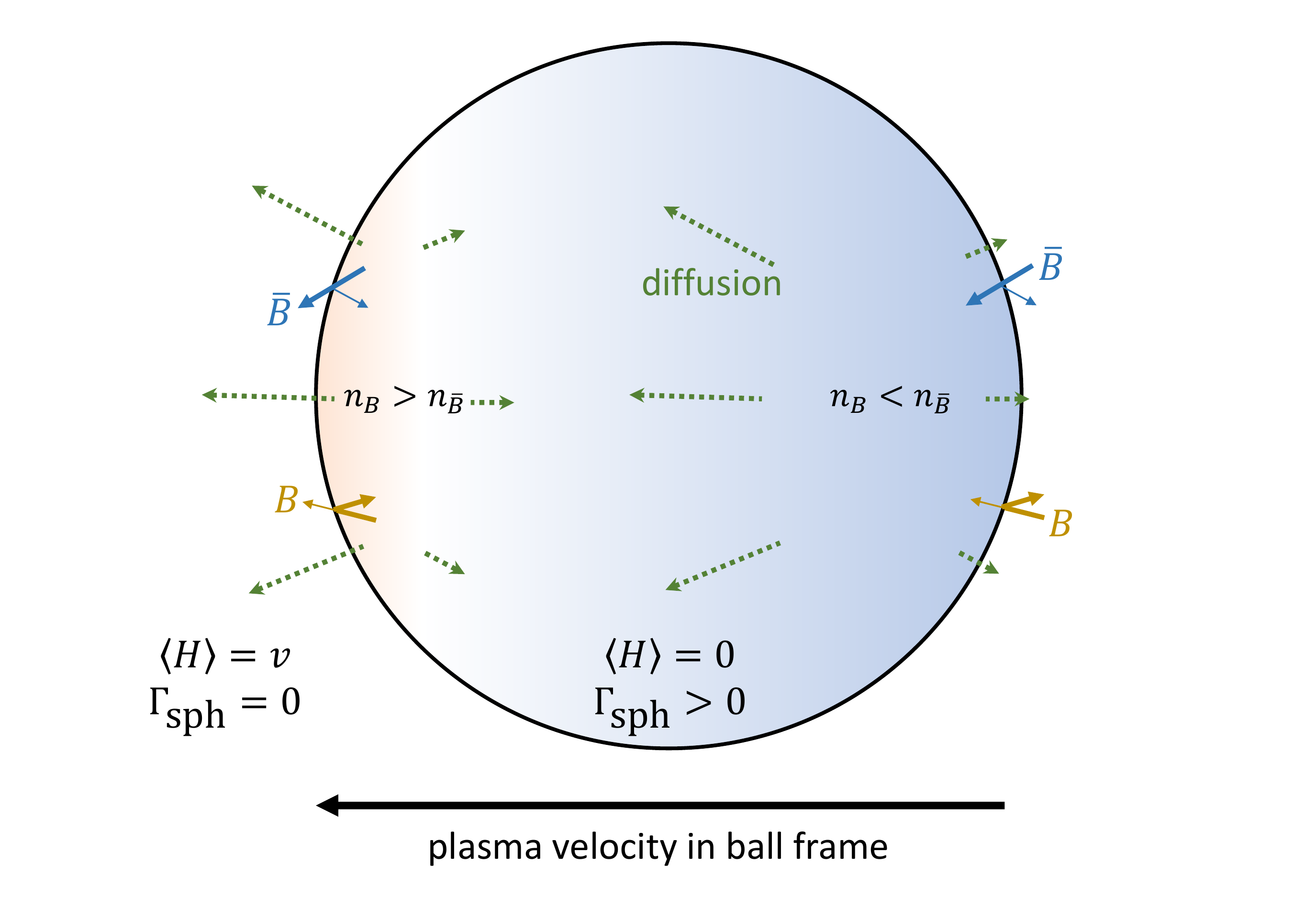}
		\caption{Schematic depiction of catalyzed baryogenesis facilitated by an electroweak symmetric ball. As the ball moves through the plasma, a baryon asymmetry builds up near its walls due to a CP-violating interaction at the bubble wall. This relies on the relative plasma-ball velocity so that more antibaryons (blue) and baryons (orange) approach the walls from the right side than from the left. The asymmetry then diffuses away from the regions near the wall, but because of the relative velocity between the ball and the plasma, particles near the leading edge tend to diffuse inwards while particles near the trailing edge tend to diffuse outwards (indicated by the green dotted lines). This creates a net baryon number asymmetry inside the ball. Sphalerons, which are only active inside the ball, then convert some of the baryons into leptons, preserving the asymmetry even as the ball passes by.		}
	\end{center}
\end{figure}

To illustrate this framework, we explore in this work the case where electroweak-symmetric (EWS) balls act as the catalyst. Such states can arise in many theories, including non-topological solitons \cite{Khlebnikov:1986ky,Dvali:1997qv,Ponton:2019hux}, dark monopoles \cite{Bai:2020ttp}, or magnetically charged black holes \cite{Lee:1991qs,Lee:1991vy,Maldacena:2020skw,Bai:2020spd,Bai:2020ezy}.
Schematically, the catalyzed baryogenesis process for this particular catalyst is illustrated in Fig.~\ref{fig:catalyze-schematic}. Inside the catalyst region, electroweak sphalerons are active because electroweak symmetry is restored, allowing baryons to be converted to leptons, while outside sphaleron-mediated processes are suppressed, providing condition (i). Sphalerons also violate C, partially satisfying condition (ii).  Interactions between the EWS ball wall and incoming particles can violate CP, completing condition (ii). Finally, the process locally goes out of equilibrium (iii) when there is a relative velocity between the EWS ball and the surrounding plasma. Due to the relative velocity, more plasma particles are incident on the right side of the ball walls than the left in Fig.~\ref{fig:catalyze-schematic}. Then, the excess antibaryons transmitted through the leading wall tend to diffuse into the ball where sphalerons are active; meanwhile, the excess baryons reflected by the trailing wall tend to diffuse out of the ball, where the sphaleron rate is suppressed following the electroweak phase transition (EWPT). Thus, the excess antibaryons inside the ball are depleted, while the excess baryons outside the ball are not, leading to a net asymmetry after the ball has passed. To satisfy condition (iii), the CP asymmetry produced at the wall should have time to diffuse sufficiently widely in the ball to have sphalerons act on the CP asymmetry before it decays via C-violating, but CP-conserving, chirality-flipping interactions. 
Furthermore, we will assume that the system remains in a dynamical equilibrium, with the plasma inside the ball quickly diffusing as the ball slowly changes direction by scattering off the plasma.  This equilibrium situation is illustrated in Fig.~\ref{fig:catalyze-schematic}.  The picture breaks down if the ball changes direction sufficiently quickly that the plasma inside the ball does not have time to return to this dynamical equilibrium between order one changes in direction.

CAB has some similarities to the electroweak baryogenesis model based on a strongly first-order EWPT. For both cases, the motion of the bubble or ball wall provides the out-of-equilibrium condition. Different from CAB, electroweak baryogenesis has the outside region active for the sphaleron process. Furthermore, the bubbles expand, collide, percolate, and disappear shortly after generating baryon asymmetry. On the contrary, the EWS balls in CAB could survive for a cosmologically long time. If they survive until the present day, they could be searched for directly. Otherwise, if they evaporate, the resulting particles can leave cosmological imprints.
Further, CAB does not rely on any assumptions about the nature of the EWPT---excepting perhaps the temperature at which it occurs. In CAB, baryogenesis is not driven by a phase transition.

Because the catalysts considered here are EWS, interactions with Standard Model (SM) particles are unavoidable, leading to large interactions between the balls and nuclei~\cite{Ponton:2019hux,Bai:2019ogh}. One important finding of this work is that while EWS balls could explain both baryogenesis and dark matter simultaneously, they would have already been observed by direct detection experiments. Therefore, EWS balls of a given mass and radius may either play the role of dark matter or catalyst for baryogenesis, but not both. To be a catalyst, they should decay or evaporate into SM particles or dark-sector particles in the early universe. This opens a narrow region of parameter space where a symmetric population of EWS balls could catalyze baryogenesis. If the EWS balls are assumed to themselves carry an asymmetric component, the parameter space significantly expands. In the last section, we also speculate on other possible catalyst models that may have suppressed direct detection cross sections, forgoing the need for their eventual evaporation.

Similar scenarios were previously studied in Refs.~\cite{Brandenberger:1994mq,Cline:1998rc}, with a particular focus on EWS cosmic strings instead of balls. Some of our detailed analysis follows the treatment from these works. For both ball and string scenarios, the final baryon asymmetry is found to be proportional to the catalyst velocity square. However, the strings they consider only have radius of order $v^{-1}$ with $v=246$~GeV the Higgs field vacuum expectation value (VEV), whereas we consider EWS balls with much larger radius. Further, baryogenesis is limited in their case by the string network evolution: friction-dominated strings move too slowly, while in the scaling regime strings are too scarce. Thus, while Ref.~\cite{Cline:1998rc} reached a negative conclusion for strings to catalyze baryogenesis, our work finds that balls can explain the observed baryon asymmetry due to these differences.

This paper is organized as follows. In the next section, we describe the CAB mechanism in detail, including how all Sakharov conditions are satisfied. In Sec.~\ref{sec:abundance}, we synthesize the conditions to detail the allowed parameter space of EWS balls as catalysts of baryogenesis. The mechanisms and effects of EWS ball decays are described in Sec.~\ref{sec:evap}.  The final section in Sec.~\ref{sec:conclusions} gives concluding remarks.

\section{Catalyzed baryogenesis by an EWS ball}

This section will estimate the amount of baryon asymmetry that is produced. We begin by assuming that EWS balls exist after the electroweak phase transition, with mass $M_\ball$ and radius $R_\ball$. We do not specify the detailed constituents or initial formation mechanism of the EWS ball, although we will occasionally reference model-specific considerations for the non-topological soliton that is made of a complex scalar field $\Phi$ with a Higgs-portal coupling to the SM (see Ref.~\cite{Ponton:2019hux} for details).
We express the EWS ball abundance as a fraction of the dark matter (DM) abundance $\rho_\ball = f_\text{DM} \, \rho_\text{DM}$. It is convenient to write this in terms of the yield (with $s=\frac{2 \pi^{2}}{45}g_{*s}T^{3}$ the entropy density and $g_{*s}$ the number of effective relativistic degrees of freedom):
\beqa\label{eq:DMBDM}
Y_{\ball} = \frac{n_{\ball}}{s} \approx 5.1 \times 10^{-17} \, f_\text{DM} \left( \frac{10^7 \, \text{GeV}}{M_\ball} \right) \, .
\eeqa
Here, $n_\ball = \rho_\ball/M_\ball$ is the EWS ball number density, and $f_\text{DM}$ can be a function of temperature $T$ if the ball has some non-trivial evolution history.

\subsection{Out-of-equilibrium condition}

First, we address the out-of-equilibrium condition which creates a baryon asymmetry inside of the EWS ball. Here, we assume the existence of a CP-violating interaction at the EWS ball wall, which we later justify.

The basic mechanism at play is that there is a localized loss of equilibrium around the wall of the EWS ball as it passes through the plasma and disturbs it. In particular, as we will show, the wall can generate a localized CP-asymmetry in its wake, specifically a difference in the abundance of left-handed top quarks compared to right-handed anti-top quarks. Chirality-flipping interactions via the weak force tend to damp this asymmetry. So long as the asymmetry persists locally within part of the EWS ball, the out-of-equilibrium condition is satisfied. In the remainder of this subsection, we determine the chemical potential for this CP asymmetry using the diffusion equation.

The diffusion equation for the CP asymmetric number density of fermions in the presence of a CP-violating source is given by \cite{Joyce:1994zn,Joyce:1994zt,Cline:2000nw}
\beqa
\label{eq:diffusion_diffeq-3D}
\frac{\partial n_{_{\rm CP}} (\vec{x}, t)}{\partial t}- D\, \nabla^2 n_{_{\rm CP}}(\vec{x}, t) + \Gamma \, n_{_{\rm CP}}(\vec{x}, t) = S(\vec{x}, t)  \, .
\eeqa
Here, $D$ is the fermion diffusion constant and $\Gamma$ is the rate of chirality damping, including spin-flip interactions (for the top quark, $D\approx 6/T$ and $\Gamma \approx T/100$~\cite{Fromme:2006cm}). As the EWS ball moves in the plasma, before the ball changes directions, the incoming particle source comes from one direction. In the limit of large ball radius with $R_\ball \gg \sqrt{D/\Gamma}$, one can approximately treat the problem as a one-dimensional problem.
Provided that any change in the direction and velocity of the EWS ball rapidly leads to a quasi-static solution, we can seek a steady state solution in which there is a fixed moving source with velocity $v_{\rm w} \sim \sqrt{T/M_\ball}$ for the balls thermalized with plasma. The solution is $n_{_{\rm CP}}(x,t)=n_{_{\rm CP}}(x-v_{\rm w}\,t)$, and the equation can be reduced to a single variable $x$:
\beqa
(-v_{\rm w}\, \partial_{x} -D \, \partial_{x}^{2}) \, n_{_{\rm CP}} + \Gamma \, n_{_{\rm CP}} = S(x) \, .
\eeqa
This equation can be solved using a Green's function [the solution when $S(x)=\delta(x-x')$] given by
\begin{equation}
G(x-x') = \frac{D^{-1}}{k_{+}-k_{-}} \begin{cases} 
\exp[-k_{+}(x-x')] & \quad x>x' \\
\exp[-k_{-}(x-x')] & \quad x<x' 
\end{cases}
\, ,
\end{equation}
where,
\begin{equation}\label{eq:kpkm}
k_{\pm} = \frac{v_{\rm w}}{2 D} \left( 1 \pm \sqrt{1+\frac{4 \Gamma D}{v_{\rm w}^{2}}} \right) \, .
\end{equation}
In the limit of $v_{\rm w}^2 \ll \Gamma D$, $k_{\pm} \approx \pm \sqrt{\Gamma/D}$.

For large $R_\ball$, the ball can be approximated as two walls/sources propagating back-to-back but with opposite CP-violating sign. Thus, the source term is approximately $S_{2}(x)=S(x+R_\ball) - S(x-R_\ball)$, where the source function at each wall is given by \cite{Cline:2000nw,Cline:1998rc}
\begin{equation}
\label{eq:source}
S(x)\approx \frac{v_{\rm w}\, y_{f}^{2}\,  D\, T^{2}\,  \theta'''}{6} \, .
\end{equation}
Here $y_f$ is the fermion Yukawa coupling to the Higgs field; $\theta(x)$ is the wall profile of the CP-violating angle, which can be approximated by $\theta'(x)=\Delta \theta \, \delta(x)$ with $\Delta \theta$ as the angle changes from its outside to inside values~\cite{Cline:1998rc}. The derivation of (\ref{eq:source}) uses the WKB approximation, which is valid when the wall thickness $\sim v^{-1}$ is larger than the de Broglie wavelength of the particles $\sim (3T)^{-1}$. Because of this, we must assume
\begin{equation}
\label{eq:radius}
R_\ball \gtrsim v^{-1} \, ,
\end{equation}
which is consistent with the previous condition $R_\ball \gg \sqrt{D/\Gamma}$ to use a one-dimensional diffusion equation. With this source term, the CP asymmetric number density is given by
\beqa
\label{eq:CP-number-density}
n_{_{\rm CP}}(x) &=& \int_{-\infty}^{\infty}G(x-x') S_{2}(x') dx'  \nonumber \\
&=& \frac{v_{\rm w} \, y_{f}^{2} \, T^{2} \Delta \theta}{6(k_{+}-k_{-})} \left( k_{+}^{2} \, \exp[-k_{+}(x+R_\ball)]-k_{-}^{2} \, \exp[-k_{-}(x-R_\ball)] \right) \,.
\eeqa
For the EWS ball, the region of interest is $x \in (-R_\ball,R_\ball)$, where the sphalerons are active.
In the limit of $v_{\rm w} \ll \sqrt{\Gamma D}$ and $|k_{-}R_\ball|\approx|k_{+}R_\ball| \gg 1$, the radius-averaged number density inside the ball is 
\beqa
\overline{n}_{_{\rm CP}} = \frac{\int_{-R_\ball}^{R_\ball} n_{_{\rm CP}}(x) dx}{2R_\ball} \approx 
 \frac{v_{\rm w}^{2}\, y_{f}^{2}\, T^{2}\,\Delta \theta}{24 \, R_\ball \, \sqrt{\Gamma\, D}} \, .
\eeqa
Notice that due to the cancellation of the contributions from the two walls in (\ref{eq:CP-number-density}), the average number density is proportional to the square of the wall velocity. Also, from \eqref{eq:CP-number-density}, the number density is localized around a distance $1/|k_\pm| \approx \sqrt{D/\Gamma}$ within the ball wall. This is larger than the thickness of the wall, justifying the above approximation of $\theta(x)$ as a step function for the integral in (\ref{eq:CP-number-density}).

The CP-violating chemical potential can then be evaluated by $\mu_{\rm CP} = 2\,\overline{n}_{_{\rm CP}}/T^{2}$~\cite{Cline:2006ts}. For the top quark, $\Gamma \approx T/100$, $D\approx 6/T$, and $y_{f} \approx 1$~\cite{Fromme:2006cm}, giving 
\beqa
\mu_{\text{CP}} \approx  \frac{v_{\rm w}^{2} \,  \Delta \theta}{R_\ball} \, .
\eeqa

The EWS balls are not actually moving in one direction  because of their interactions with SM particles in the plasma. 
The ball thermalization rate $\Gamma_{\text{therm}} \simeq \sum_i \left(n_{i}\, \sigma_i v_{\text{rel},i} \right) T / M_\ball$ is related to the EWS ball's cross section $\sigma_i \simeq \pi R_\ball^2 \, |\mathcal{R}_i|^2$ with SM particles in the plasma (indexed by $i$) whose number density is $n_i$. The reflection probability $|\mathcal{R}_i|^2 \sim 1$ for semi-relativistic or non-relativistic species, while $n_i$ is suppressed for non-relativistic species. Thus, $\Gamma_{\text{therm}} \sim  g\, T^{4}R_\ball^{2}/M_\ball$, with $g$ the number of semi-relativistic degrees of freedom in the bath. To justify our previous estimation for $\overline{n}_{_{\rm CP}}$, the thermalization rate is required to be smaller than the diffusion rate, or $\Gamma_{\text{therm}} \lesssim D\,|k_\pm|^2 \approx \Gamma$, which gives
\beqa
\label{eq:thermalization-condition}
M_\ball \gtrsim \frac{g\,T^4}{\Gamma}R_\ball^2 \, \approx \,  10^3\, T^3\,R_\ball^2 ~,
\eeqa
where $g \approx 9$, taken to be the $W$ and $Z$ degrees of freedom assuming CAB takes place shortly after the EWPT.
When the thermalization rate is very high in violation of \eqref{eq:thermalization-condition}, non-steady solutions should be obtained to estimate the $\overline{n}_{_{\rm CP}}$. In the ball rest-frame, the source $S(\vec{x}, t)$ in (\ref{eq:diffusion_diffeq-3D}) should vary in time intervals of order $1/\Gamma_{\text{therm}}$. We do not calculate $\overline{n}_{_{\rm CP}}$ for this case in this work and will restrict the model parameter space to satisfy \eqref{eq:thermalization-condition}. As we will later see in Eq.~(\ref{eq:massradius-bound}), the region where (\ref{eq:thermalization-condition}) is not satisfied is anyways mostly excluded by other unrelated considerations.

\subsection{Baryon number violation}

With the chemical potential inside the EWS ball derived in the previous section, a net baryon asymmetry can be produced via baryon number violating sphalerons. When the ball moves slowly, the EW sphaleron can convert the CP asymmetry into baryon asymmetry via the ``non-local" way~\cite{Joyce:1994zt}. 

The baryon number generated per unit time by a single EWS ball is given by \cite{Cline:1998rc} 
\beqa
\frac{dN_{B}}{dt} 
\approx - \Gamma_{\text{sph}}\, R_\ball^{3}\, \frac{\mu_{\text{CP}}}{T} \, ,
\eeqa
where $\Gamma_{\text{sph}}\approx 25\, \alpha_{w}^{5}\,T^{4}\,\approx 10^{-6} \,T^{4}$ is the sphaleron rate per unit volume \cite{Cline:2006ts}.  Here, we have used that $\Gamma_{\text{sph}} / T^3 < |k_\pm|$, implying particles undergo less than one sphaleron transition on average before the asymmetric region near the wall diffuses. The density of produced baryon asymmetry $n_B$ is related to the EWS ball density by 
\begin{equation}
\frac{dn_{B}}{dt} + 3 H n_{B}= - \Gamma_{\text{sph}} R_\ball^{3} \frac{\mu_{\text{CP}}}{T} n_\ball \, .
\end{equation}
Using $\frac{dn_{B}}{dt} + 3 H n_{B} = s \,dY_{B}/dt$ with $s$ as the entropy density and $dT/dt= - H\, T$, 
\begin{equation}\label{eq:assymetry}
\frac{dY_{B}}{dT} = \frac{\Gamma_{\text{sph}}\,R_\ball^{3}}{H\,T}\,\frac{\mu_{\text{CP}}}{T} Y_\ball \, .
\end{equation}
Using $H(T) = 16.6 \times (g_{*}/100)^{1/2} T^{2}/M_{\rm pl}$ (assuming radiation domination), this equation can be integrated from an initial temperature $T_i$ to obtain (assuming constant $Y_\ball$)
\begin{equation} \label{eq:finalasym}
Y_{B} = 1.9 \times 10^{-10} \, f_\text{DM} \left(\frac{R_\ball}{1 \, \text{GeV}^{-1}}\right)^{2} \left(\frac{10^{8}\, \text{GeV}}{M_\ball} \right)^{2} \left(\frac{\Delta \theta}{-1} \right) \left(\frac{T_i}{100 \, \text{GeV}} \right)^{2} \, .
\end{equation}
In general, $T_i$ will be the smaller of either the EW phase transition temperature $T_\text{EWPT}$ (so that the Higgs has a nonzero VEV outside the EWS ball) or the EWS ball formation temperature $T_\text{form}$. This quantity is related to the baryon-to-photon ratio by $Y_B = \eta/7.04 \approx 0.85\times 10^{-10}$ to explain the observed baryon asymmetry.

\subsection{CP violation} 
\label{sec:CPV}

CP violation can be introduced via a simple effective operator coupling the EWS ball to the SM. For example, the constituent scalar field $\Phi$ of the EWS ball can have a CP-violating interaction with the top quark \cite{Cline:2012hg} given by 
\begin{equation}\label{eq:cpv-operator}
\mathcal{L} \supset y_{t}\,\overline{Q}_{L} \widetilde{H} \left(1 + \eta \frac{\Phi \Phi^\dagger}{\Lambda^{2}}\right) \, t_{R} + h.c. ~,
\end{equation}
where $\widetilde{H}= i\sigma_2 H^*$ and  $\eta$ is a complex parameter taken to be $e^{i\pi/2}$. If $|\Phi| \equiv \phi(r)$ varies inside and outside the ball, this induces a spatially varying complex mass for the top quark,
\begin{equation}
m_{t}(r) = \frac{y_{t}}{\sqrt{2}} \, h(r) \, \left(1 + i \frac{\phi(r)^{2}}{\Lambda^{2}} \right) = |m_{t}(r)|e^{i\theta(r)} \, ,
\end{equation}
where $\tan{\theta(r)}=\phi(r)^{2}/\Lambda^{2}$. Outside the EWS ball, $\phi(r)$ is constant, and the phase can be absorbed by a redefinition of fields. Near the EWS ball, the change in phase is physical, inducing a CP-violating angle $|\Delta\theta| \approx f^{2}/\Lambda^{2}$ for $f \lesssim \Lambda$, where $f$ is the change in the $\phi(r)$ between the interior and exterior of the ball.

For this model to provide an EWS ball, there must be a Higgs portal coupling to the $\Phi$ field (see, \eg, \cite{Ponton:2019hux,Bai:2020ttp}) with $\mathcal{L}_{\text{portal}} = \lambda \, \Phi \Phi^\dagger H H^\dagger$  with $|\lambda| f^2 \gtrsim v^2$. 
The sign of $\lambda$ depends on whether $\phi(r)^2$ increases or decreases inside the ball, which will differ depending on the model. Then, this portal coupling in conjunction with the interaction term in \eqref{eq:cpv-operator} leads via a $\Phi$ loop diagram to a dimension-6 CP-violating operator involving only SM particles, 
\beqa
\mathcal{L}_{6} \sim i \frac{ y_{t}\, \lambda }{16 \pi^{2}} \frac{|H|^{2}}{\Lambda^{2}}\overline{Q}_{L} \widetilde{H}  t_{R} \,+\, h.c.
\eeqa
The coefficient of this operator is constrained by the current upper limit on the electron dipole moment $|d_{e}|<1.1 \times 10^{-29} \, e \,\text{cm}$ by the ACME collaboration~\cite{Andreev:2018ayy}. Thus, $\Lambda > v\, (\lambda/0.13)^{1/2}$~\cite{Fuchs:2020uoc}.
If $\Lambda \gtrsim f \gtrsim v$, the EWS ball can simultaneously satisfy $|\lambda| f^2 \gtrsim v^2$, $|\Delta \theta| \lesssim 1$, and the ACME constraint. In other words, the scale of the new physics need not be much larger than the electroweak scale while still allowing $\cO(1)$ CP violation at the EWS ball wall.

\section{Abundance of EWS balls and their evolution}
\label{sec:abundance}

We now turn our attention to the parameter space where EWS balls can facilitate baryogenesis. We will require throughout that the mass-radius relation of the EWS balls satisfies 
\beqa
\label{eq:massradius-bound}
M_\ball \gtrsim \frac{4\pi}{3} v^{4} R_\ball^{3} + 4 \pi v^3 R_\ball^2 \, .
\eeqa
This represents the minimum mass contribution from the Higgs field, assuming the difference in the vacuum energy between the EWS region and the normal vacuum $\Delta V \gtrsim v^4$ for the first term. Depending on the EWS ball model, additional contributions to the mass could come from other fields or matter (\eg, from the $\Phi$ field for a Q-ball, the dark Higgs and dark gauge fields for a dark monopole, or the black hole in a magnetically charged black hole). We saw in Sec.~\ref{sec:CPV} that the scale of the EWS ball physics could be near the weak scale or much larger. Thus, these additional mass contributions may be of the same order as (\ref{eq:massradius-bound}) or much larger, making this a generic lower bound.
As an aside, for some models, it is possible to have a fine-tuned cancellation between terms so that $\Delta V \approx 0$, suppressing the first term of (\ref{eq:massradius-bound}). In this case, the gradient energy of the fields near the ball wall provides the irreducible contribution from the second term. We will not consider this tuned scenario in what follows, and it would anyways not affect our conclusions.

\subsection{The EWS ball as dark matter}

As a simplest first assumption, we can take the EWS ball yield $Y_\ball$ to be constant before the EWPT, with its abundance determined at some higher temperature scale. Then, the baryon asymmetry is fixed by Eq.~(\ref{eq:finalasym}). It is maximized when $f_\text{DM}=1$, in which case the EWS ball also plays the role of dark matter.

\begin{figure}[t!]
	\centering
\includegraphics[width=0.6\textwidth]{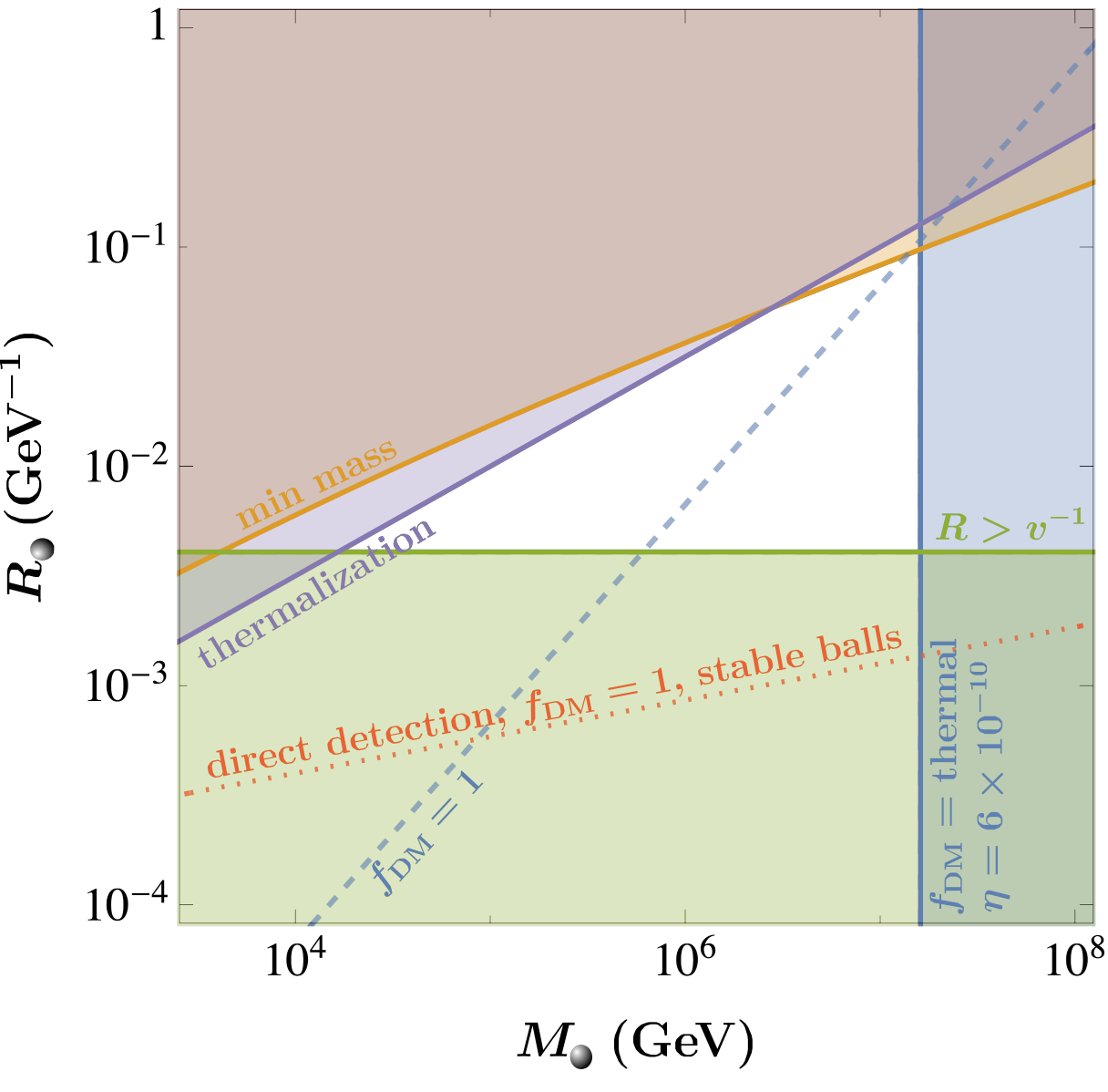}
	\caption{Bounds on the parameter space for EWS balls to catalyze baryogenesis. Blue curves show the constraints from requiring sufficient baryogenesis to explain the observed asymmetry. The solid blue curve shows the constraint assuming the EWS ball abundance is set by annihilations and includes an enhancement factor $\xi(y)=10$ by assuming annihilations are taking place during baryogenesis and $T_\text{form}=T_i$; see (\ref{eq:asym-annihilation}). The dashed blue curve shows the constraint if $f_\text{DM}$ is fixed to unity, with no enhancement factor for annihilations; see (\ref{eq:finalasym}). Both take $T_i=100 \, \GeV$ and $|\Delta \theta| \leq 1$. The orange curve enforces the mass-radius relation in (\ref{eq:massradius-bound}), while the green limits $R>v^{-1}$ from (\ref{eq:radius}). Purple enforces that the thermalization time must be long enough according to (\ref{eq:thermalization-condition}). Finally, the red dotted line shows the direct detection limits from Xenon1T \cite{Aprile:2018dbl}. It assumes $f_\text{DM}=1$, but it does not impose any constraint if the EWS balls decay after catalyzing baryogenesis.}
	\label{fig:bounds_annihilation_decay}
\end{figure}

The parameter space where EWS balls can catalyze baryogenesis is shown in Fig.~\ref{fig:bounds_annihilation_decay}, setting $f_\text{DM}=1$. The conditions of producing enough asymmetry in (\ref{eq:finalasym}) (dashed blue line, excluding the region to its right by demanding $|\Delta \theta| \leq 1$ and $T_i < T_\text{EWPT} \sim 100~\GeV$~\cite{Laine:2015kra,DOnofrio:2015gop}), having a long enough thermalization time in (\ref{eq:thermalization-condition}) (purple), satisfying the mass condition (\ref{eq:massradius-bound}) (orange), and having a large enough radius in (\ref{eq:radius}) (green) are imposed. 
It is clear that these conditions allow for successful baryogenesis.  

However, the relic abundance of EWS balls can be probed at direct detection experiments. 
Following \cite{Ponton:2019hux,Bai:2019ogh,Bai:2020ttp}, we approximate the EWS ball-nucleus scattering cross section as the smaller of the expectation from the Born approximation and the geometric cross section, 
\begin{equation}
\label{eq:ddxsec}
\sigma_{\ball A} \approx \min \left[\frac{16\pi}{9} m_N^2 A^4 y_{hNN}^2 v^2 R_\ball^6, \, 2\pi R_\ball^2\right] \, , 
\end{equation}
where $m_N$ is the nucleon mass, $A$ is the nucleus's atomic mass number, and $y_{hNN}\approx 1.1\times 10^{-3}$.  Direct detection experiments like Xenon1T \cite{Aprile:2018dbl} can then place upper bounds on surviving EWS ball radii as shown by the dotted red curve in Fig.~\ref{fig:bounds_annihilation_decay}.
Due to the combination of constraints, it is clear that it is not possible to explain both baryogenesis and dark matter simultaneously with EWS balls. Further, even if $f_\text{DM}$ is reduced, the direct detection constraint on $R_\ball$ will weaken only by $f_\text{DM}^{-1/6}$ [see Eq.~(\ref{eq:ddxsec})] while the baryon asymmetry is proportional to $f_\text{DM}$, leaving no available parameter space that satisfies all constraints.

\subsection{EWS ball annihilations}

Clearly, a fixed value for $f_\text{DM}$ is constrained by the competition between direct detection constraints and creating a large enough baryon asymmetry. Instead, one could consider that $f_\text{DM}$ evolves over time. If the EWS ball abundance is larger near $T \sim 100~\GeV$ than it is today, then baryogenesis could still be sufficiently efficient while evading direct detection constraints. In this subsection we will show that $f_\text{DM}$ can evolve due to EWS ball annihilations, but the impact on baryogenesis is relatively small.

Here, consider a purely symmetric population of EWS balls. Further, if the EWS ball is a non-topological soliton, assume that most of the $\Phi$ particles are bound inside EWS balls. If an initial population of EWS balls form at some initial temperature $T_\text{form}$ with abundance $Y_\text{form}$, then some of them have a chance to annihilate. For simplicity, assume all EWS balls have the same mass and radius. Further assume that EWS balls of opposite charges will annihilate with geometric cross section $\sigma \sim \pi R_\ball^2$ (\ie, there is no long range force, just contact interactions). Then the evolution of EWS balls will go as 
\begin{align}
\frac{dn_\ball}{dt} + 3 H n_\ball = -\langle\sigma v_\text{rel}\rangle n_\ball^{2} \qquad \Rightarrow \qquad  \frac{dY_\ball}{dT} = \frac{\langle\sigma v_\text{rel}\rangle\,s\, Y_\ball^{2}}{H\,T} ~.
\end{align} 
Taking $g_{*} = g_{*s} \sim 100$ and $v_\text{rel}=(T/M_\ball)^{1/2}$,
\begin{equation}
\label{eq:annihilation-solution}
Y_\ball(T)^{-1} - Y_\ball(T_\text{form})^{-1} = R_\ball^{2} \, M_\ball^{-1/2}\,g_{*}^{1/2}\,M_{\text{pl}}\, (T_\text{form}^{3/2}-T^{3/2}) \, .
\end{equation}

For this to make a difference in $Y_B$, $T_\text{form}$ must be close to the EWPT temperature $T_\text{EWPT}$. The reason: if $T_\text{form}$ is much larger, then $Y_\ball(T \ll T_\text{form})$ is constant, reducing to the previous case of constant $f_\text{DM}$; on the other hand, if $T_\text{form} \geq T_i$ is much smaller than $T_\text{EWPT}$, the baryon asymmetry is suppressed, going as $T_i^2$ in (\ref{eq:finalasym}). If we tune $T_\text{form}\sim T_\text{EWPT}$, we can plug $Y_\ball(T)$ into (\ref{eq:assymetry}) to obtain the final baryon asymmetry $Y_B$. This gives an enhancement factor to the right side of Eq.~(\ref{eq:finalasym}): 
\begin{equation}
\label{eq:enhancement}
\xi(y) = \frac{4}{3} \log (y) \, , \; \; \; \; \; \; \text{for} \; \; y \equiv \frac{Y_\ball(T_\text{form})}{Y_\ball(T \ll T_\text{form})} \gg 1 \, ,
\end{equation}
assuming $T_\text{form}=T_i < T_\text{EWPT}$, where $f_\text{DM}$ in (\ref{eq:finalasym}) would be calculated using $Y_\ball(T \ll T_\text{form})$.
At most, we expect $\xi(y)$ to give an $\cO(10)$ enhancement due to the logarithmic dependence on $y$.

Thus, including the time evolution of the EWS balls due to annihilations does not provide a large enough baryogenesis enhancement to escape direct detection bounds. It is worth noting that EWS ball annihilations were assumed instantaneous and complete in this calculation. If the annihilations were to take a long enough time, \eg, because the EWS balls form metastable bound states before annihilating, then baryogenesis could proceed with a larger EWS ball abundance than assumed here.

\subsection{EWS ball decay}

Another possibility for time dependence is to allow the EWS balls to decay. Then, they could exist shortly after the EWPT, but decay long before today, evading direct detection constraints. They may even have an abundance larger than dark matter in the early universe.

The EWS ball abundance after annihilations in the limit $Y_\ball(T_\text{form}) \gg Y_\ball(T)$ and $T_\text{form} \gg T$ from (\ref{eq:annihilation-solution}) is,
\begin{equation}
\label{eq:ball-annihilation-abundance}
Y_\ball(T \ll T_\text{form}) = 10^{-18} \left(\frac{M_\ball}{10^{7}\, \text{GeV}} \right)^{1/2} \left(\frac{0.1 \, \text{GeV}}{R_\ball} \right)^{2} \left(\frac{200 \, \text{GeV}}{T_\text{form}} \right)^{3/2} \, .
\end{equation}
Then, using \eqref{eq:DMBDM}, \eqref{eq:finalasym}, (\ref{eq:enhancement}), and (\ref{eq:ball-annihilation-abundance}),
\begin{equation}
\label{eq:asym-annihilation}
Y_{B} = 3.8 \times 10^{-11} \Big{(}\frac{10^{7} \, \text{GeV}}{M_\ball} \Big{)}^{1/2} \Big{(}\frac{\Delta \theta}{-1}\Big{)} \Big{(}\frac{200 \, \text{GeV}}{T_\text{form}} \Big{)}^{3/2} \left(\frac{T_i}{100 \, \text{GeV}} \right)^{2} \left(\frac{\xi(y)}{10}\right) \, .
\end{equation}
Thus, the necessary baryon asymmetry could be generated for $M_\ball \lesssim 10^7~\GeV$. This is shown in Fig.~\ref{fig:bounds_annihilation_decay} by the solid blue vertical line (the dotted red direct detection constraint and dashed blue $f_\text{DM}=1$ constraint can be ignored here, since the abundance is assumed set by annihilations and decays well before today). Note that $T_\text{form}$ cannot be much larger than $T_\text{EWPT}$, otherwise not enough baryon asymmetry is produced (in conjunction with lower bounds on $M_\ball$ in Fig.~\ref{fig:bounds_annihilation_decay}). Also, since $T_i \leq T_\text{form}$, the formation temperature cannot be too small either. In all cases, $T_i \leq T_\text{EWPT}$ is required.

\subsection{Asymmetric EWS balls}

If the dark sector responsible for EWS ball formation itself carries an asymmetry, then the EWS ball annihilations will halt once the symmetric component is depleted. This presents advantages compared to the prior symmetric case. First, it could allow $Y_\ball$ well in excess of (\ref{eq:ball-annihilation-abundance}), enhancing the amount of baryogenesis. Second, it removes any dependence for $Y_B$ on $T_\text{form}$, as long as $T_\text{form} \geq T_\text{EWPT}$. Third, a model-dependent advantage for the specific case of EWS Q-balls is that free $\Phi$ antiparticles are also depleted, making it easier for Q-balls to survive without assuming every $\Phi$ particle must be bound in a Q-ball \cite{Griest:1989bq}. For the EWS ball abundance to be compatible with baryogenesis, it would be constrained by direct detection at low mass and overclosure at high mass if the EWS balls were stable. Thus, this possibility still depends on the EWS balls being unstable and evaporating between the EWPT and today. 

\begin{figure}[t!]
	\centering
  	\includegraphics[width=0.6\textwidth]{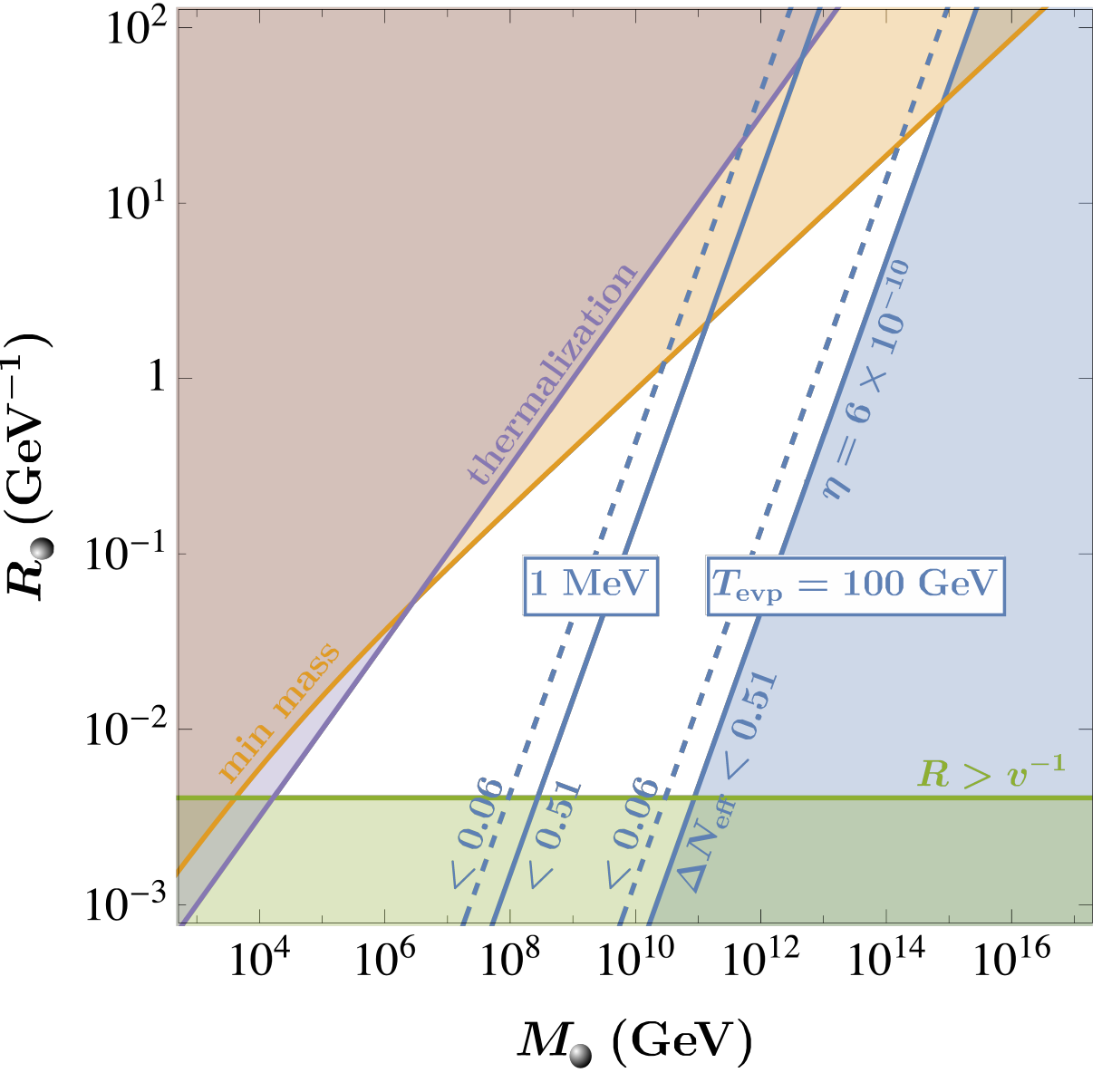}
	\caption{Same as Fig.~\ref{fig:bounds_annihilation_decay}, but now allowing an asymmetric component of EWS balls that eventually decay into massless states in the dark sector. Constraints from $\Delta N_\text{eff}$ limit where baryogenesis can produce large enough $\eta$. Examples for present (projected) constraints are shown in solid (dashed) blue lines for two different possible evaporation temperatures. Because $T_\text{evp}<T_\text{EWPT}\sim 100~\GeV$ is required for baryogenesis by EWS balls, the shaded blue region on the right is completely excluded by current constraints. If the dark sector states are massive, additional constraints can apply as discussed in the text.}
	\label{fig:bounds_asymmetric_decay}
\end{figure}

The asymmetric abundance is constrained by requiring the decay products of the EWS balls not contribute too much to the effective radiation degrees of freedom during BBN or recombination.  The most conservative bound comes from assuming EWS balls decay shortly after the EWPT. As shown in the next section, this amounts to the requirement $f_\text{DM} \lesssim 2 \times 10^{10}$ in Eq.~(\ref{eq:finalasym}). Clearly, this opens a large swath of parameter space in $R_\ball$, $M_\ball$, and $\Delta \theta$ that is not accessible when the EWS-ball-forming sector is symmetric, which is depicted in Fig.~\ref{fig:bounds_asymmetric_decay}.

\section{Evaporation of EWS balls}
\label{sec:evap}

Supposing that the EWS ball can evaporate, we denote the rate of evaporation by $\Gamma_{\text{evp}}$. Such evaporation can happen, for example, by directly decaying the field $\Phi$ making up the EWS ball into new dark sector particles. For instance, if there is a $U(1)_{\Phi}$-breaking coupling of the form $\mathcal{L}_{\Phi \chi} = g_{\Phi \chi}\Phi \bar{\chi} \chi + h.c.$, then evaporation can happen via decaying into $\chi$ particles, which we take to be massless for simpicity, each carrying energy  of $M_\ball/(2 Q)$ with $Q$ as the total $U(1)_\Phi$ charge of an EWS ball. The massless products of this decay then redshift as radiation. The energy contained in the $\chi$ particles will contribute to the effective number of relativistic degrees of freedom, $\Delta N_{\text{eff}}$, which is constrained by probes of BBN and CMB measurements. Alternatively, if the EWS balls decay into SM particles before BBN ($T_{\rm evp} \gtrsim 1$~MeV), there is no constraint from $\Delta N_{\text{eff}}$.

  We will take a model-independent approach to evaporation in which the temperature at which EWS balls evaporate ($T_{\text{evp}}$), {\it i.e.}, the temperature at which $H(T_{\text{evp}})=\Gamma_{\text{evp}}$, is a phenomenological parameter. The total energy density contained in EWS balls is $f_{\text{DM}}\,\rho_{\text{DM}}(T)$ which redshifts as matter from $T=100 \, \text{GeV}$ to $T=T_{\text{evp}}$. Hence the total energy density that is converted to $\chi$-particles is given by
  \begin{equation}
  \rho_{\chi}(T=T_{\text{evp}}) = 0.14 \left(\frac{f_{\text{DM}}}{10^{7}}\right) \left(\frac{g_{*s}(T_{\text{evp}})}{60}\right) \left(\frac{T_{\text{evp}}}{1\, \text{GeV}}\right)^{3} \, \text{GeV}^{4}.
  \end{equation}  
The contribution of $\rho_{\chi}$ to $\Delta N_{\text{eff}}$ at late times that the CMB is sensitive to is given by  
\begin{equation}
\Delta N_{\text{eff}}  \approx \left(\frac{8}{7}\right) \left(\frac{11}{4}\right)^{4/3} \frac{\rho_{\chi}(T_{\text{CMB}})}{\rho_{\gamma}(T_{\text{CMB}})}.
\end{equation}
Since the radiation produced in the evaporation of the EWS balls is decoupled from the SM bath, their density scales as $\rho_\chi \propto 1/a^4 \propto g_{*s}^{4/3}(T) \, T^4$. From this, we determine the contribution to the energy density at the temperature of recombination and find that
\begin{equation}
\Delta N_{\text{eff}}  \approx 0.025 \, \left(\frac{f_{\text{DM}}}{10^{7}}\right) \left(\frac{T_{\text{evp}}}{1\, \text{GeV}}\right)^{-1} \left(\frac{60}{g_{*s}(T_{\text{evp}})}\right)^{1/3}~. 
\end{equation}
Large $f_{\text{DM}}$, corresponding to more energy density in EWS balls, and small $T_{\text{evp}}$, corresponding to the late decay of EWS balls, are more constrained. The current strongest limit comes from the CMB epoch given by the Planck 2018 observations $\Delta N_{\text{eff}}< 0.51$~\cite{Aghanim:2018eyx,Riess_2018}. 
For successful baryogenesis, $T_\text{evp} \lesssim T_\text{EWPT} \sim 100 \, \GeV$ is required, setting a conservative bound $f_\text{DM} \lesssim 2 \times 10^{10}$. Next-generation observations from CMB-S4 experiments are projected to improve sensitivity by an order of magnitude to $\sigma( N_{\text{eff}})< 0.03$~\cite{Abazajian:2016yjj}. 

Alternatively, if the $\chi$ have mass $m_\chi>0$, they could explain dark matter provided $m_\chi \sim M_\ball / (2 Q f_\text{DM})$. They must not be too light, otherwise they will free stream for too long and suppress structure formation. They become non-relativistic at the time the universe has reached temperature $T_\text{NR} \sim T_{\rm evp}/f_{\rm DM}$. Their free-streaming length up to matter-radiation equality (at temperature $T_\text{eq}$), when perturbations become Jeans unstable, is $\lambda_{\rm FS} \sim \frac{2 \, M_{\rm pl}}{T_{0}\,T_{\rm NR}} \left(1+\log[\frac{T_{\rm NR}}{T_{\rm eq}}]\right)$~\cite{Kolb:1990vq}, where $T_0=2.35 \times 10^{-4} \, \text{eV}$ is the temperature today. The bound on $T_\text{NR}$ is similar to the bound on the thermal warm dark matter mass: $T_{\rm NR} \gtrsim 1$~keV~\cite{Viel:2013fqw,Hsueh:2019ynk,Gilman:2019nap,Banik:2019smi}, corresponding to $\lambda_\text{FS} \lesssim 1\,\text{Mpc}$. This sets a bound $f_\text{DM} \lesssim 10^8\times (T_\text{evp}/100\,\GeV)$, a bit stronger than the $\Delta N_\text{eff}$ constraint for this case. If $m_\chi$ is smaller than the quantity above, $\chi$ only makes up a subdominant component of dark matter, and the free streaming constraint is correspondingly relaxed.

\section{Discussion and conclusions}
\label{sec:conclusions}

We have used the EWS ball as a representative model to implement the more general catalyzed baryogenesis mechanism. One could consider other possible models to introduce baryon number violating interactions. For instance, the constituents of the catalyst ball can interact with SM particles via baryon number violating higher-dimensional operators. If those operators mainly contain the third or second-generation of quarks, the direct detection constraints could be dramatically relaxed and the catalyst balls can exist in the current universe and contribute to a significant fraction of dark matter. Similarly, lepton-number violating operators can also be adopted to catalyze leptogenesis before the electroweak phase transition. The generated lepton asymmetry is then converted into a baryon asymmetry by the electroweak sphaleron process. Additionally, one could use the catalyst mechanism to generate an asymmetry for a new particle that carries both baryon and dark matter number. The later decays of this new particle can provide a unified origin for baryon and dark matter asymmetry~\cite{Davoudiasl:2010am}.  

In summary, we have proposed a novel mechanism to generate the baryon asymmetry that is similar to the catalytic reaction in chemistry. To generate enough baryon asymmetry, the catalyst balls are preferred to have a smaller mass and a larger radius. We have used the electroweak-symmetric ball to guide our discussion of this general catalyzed baryogenesis mechanism. For the EWS ball abundance determined by their annihilations, the EWS ball mass and radius are required to be around $10^6$~GeV and $10^{-2}$~GeV$^{-1}$ to explain the observed baryon asymmetry. Interestingly, EWS ball relics within this region of parameter space are already excluded by direct detection constraints, so they are required to decay into other states in the early universe to evade the constraints. We also discussed the case with asymmetric EWS balls with a much larger initial abundance and a wider parameter space in $M_\ball$ and $R_\ball$ to accommodate the observed baryon asymmetry. The dark radiation from EWS ball evaporations provide a large contribution to the effective number of relativistic degrees of freedom, which could be tested in future CMB experiments.

\subsubsection*{Acknowledgements}
We thank Hooman Davoudiasl for insightful discussion. 
The work of YB and MK is supported by the U.S. Department of Energy under the contract DE-SC-0017647. The work of JB is supported by start up funds from Colorado State University. The work of NO is supported by the Arthur B. McDonald Canadian Astroparticle Physics Research Institute.

\setlength{\bibsep}{3pt}

\providecommand{\href}[2]{#2}\begingroup\raggedright\endgroup

\end{document}